\documentclass[a4paper,12pt]{article}
\usepackage{amsmath,amsthm}
\usepackage[letterpaper,margin=1in]{geometry}
\usepackage{blindtext}
\numberwithin{equation}{section}

\DeclareRobustCommand{\rchi}{{\mathpalette\irchi\relax}}
\newcommand{\irchi}[2]{\raisebox{\depth}{$#1\chi$}} 
\usepackage{hyperref}
\begin{document}
\centerline{}\centerline{\Large{\bf On the application of $M$-projective curvature tensor }}
\centerline{}\centerline{\Large{\bf in general relativity}}
\centerline{}\centerline{$~$Kaushik Chattopadhyay*, Arindam Bhattacharyya** and Dipankar Debnath***}
\centerline{*Department of Mathematics,}\centerline{Jadavpur University, Kolkata-700032, India}
\centerline{E-mail:kausikchatterji@gmail.com}
\centerline{**Department of Mathematics,}\centerline{Jadavpur University, Kolkata-700032, India}
\centerline{E-mail:bhattachar1968@yahoo.co.in}
\centerline{***Department of Mathematics,}\centerline{Bamanpukur High School(H.S), Nabadwip, India}
\centerline{E-mail:dipankardebnath123@hotmail.com}
\newtheorem{Theorem}{Theorem}[section]
\begin{abstract}{\em In this paper the application of the $M$-projective curvature tensor in the general theory of relativity has been studied. Firstly, we have proved that an $M$-projectively flat quasi-Einstein spacetime is of a special class with respect to an associated symmetric tensor field, followed by the theorem that a spacetime with vanishing $M$-projective curvature tensor is a spacetime of quasi-constant curvature. Then we have proved that an $M$-projectively flat quasi-Einstein spacetime is infinitesimally spatially isotropic relative to the unit timelike vector field $\xi$. In the next section we have proved that an $M$-projectively flat Ricci semi-symmetric quasi-Einstein spacetime satisfying a definite condition is an $N(\frac{2l-m}{6})$-quasi Einstein spacetime. In the last section, we  have firstly proved that an $M$-projectively flat perfect fluid spacetime with torse-forming vector field $\xi$ satisfying Einstein field equation with cosmological constant represents an inflation, then we have found out the curvature of such spacetime, followed by proving the theorem that the spacetime also becomes semi-symmetric under these conditions. Lastly, we have found out the square of the length of the Ricci tensor in this type of spacetime and also proved that if an $M$-projectively flat perfect fluid spacetime satisfying Einstein field equation with cosmological constant, with torse-forming vector field $\xi$ admits a symmetric $(0,2)$ tensor $\alpha$ parallel to $\nabla$ then either $\lambda = \frac{k}{2}(p-\sigma)$ or $\alpha$ is a constant multiple of $g$.} 
\end{abstract}
\textbf{M.S.C.2010:} ~~Primary 53C25 ; Secondary 53D10, 53C44.

\section{Introduction}

$~~~$An Einstein manifold is a Riemannian or pseudo-Riemannian manifold whose Ricci tensor S of type $(0, 2)$ is non-zero and proportional to the metric tensor. Einstein manifolds form a natural subclass of various classes of Riemannian or semi-Riemannian manifolds by a curvature condition imposed on their Ricci tensor $\cite{jmh1}$. Also in Riemannian geometry as well as in general relativity theory, the Einstein manifold plays a very important role.\\\\
$~~~~$Chaki and Maity $\cite{jmh2}$ generalised  the concept of Einstein manifold and introduced the notion of quasi-Einstein manifold. According to them, a Riemannian or semi-Riemannian manifold is said to be a quasi-Einstein manifold if its Ricci tensor S of type $(0, 2)$ is non-zero and satisfies the condition
\begin{eqnarray}\
S(U, V) = lg(U,V) + mA(U)A(V),
\end{eqnarray}
where $l$ and $m$ are two non-zero real-valued scalar functions and $A$ is a non-zero $1$-form equivalent to the unit vector field $\xi$, i.e. $g(U, \xi) = A(U)$, $g(\xi, \xi) = 1$. If $m = 0$ then the manifold becomes Einstein. Quasi-Einstein manifolds are denoted by $(QE)_{n}$, where $n$ is the dimension of the manifold. The notion of quasi-Einstein manifolds arose during the study of exact solutions of the Einstein field equations as well as during the considerations of quasi-umbilical hypersurfaces of semi-Euclidean spaces. For instance, the Robertson-Walker spacetime is a quasi-Einstein manifold. Also, quasi-Einstein manifolds can be taken as a model of perfect fluid spacetime in general relativity. The importance of quasi-Einstein spacetimes lies in the fact that 4-dimensional semi-Riemannian manifolds are related to study of general relativistic fluid spacetimes, where the unit vector field $\xi$ is taken as timelike velocity vector field, that is, $g(\xi, \xi) = - 1$.\\\\
$~~~~$In the recent papers $\cite{jmh4}$, $\cite{jmh3}$, the application of quasi-Einstein spacetime and generalised quasi-Einstein spacetime in general relativity have been studied. Many more works have been done in the spacetime of  general relativity $\cite{jmh31}$, $\cite{jmh22}$, $\cite{jmh7}$, $\cite{jmh8}$, $\cite{jmh27}$, $\cite{jmh28}$, $\cite{jmh5}$.\\\\
$~~~~$Let $(M_{n}, g)$ be an $n$-dimensional differentiable manifold of class $C^{\infty}$ with the metric tensor $g$ and the Riemannian connection $\nabla$. In 1971 G. P. Pokhariyal and R. S. Mishra $(\cite{jmh10})$ defined the $M$-projective curvature tensor as follows
\begin{eqnarray}
&&\tilde P(U,V)W = R(U,V)W - \frac{1}{2(n-1)}[S(V,W)U - S(U,W)V \nonumber\\
&&~~~~~~~~~~~~~~+ g(V,W)QU-g(U,W)QV],
\end{eqnarray}
where $R$ and $S$ are the curvature tensor and the Ricci tensor of $M_{n}$, respectively. Such a tensor field $\tilde P$ is known as the $M$-projective curvature tensor. Some authors studied the properties and applications of this tensor $\cite{jmh29}$, $\cite{jmh23}$, $\cite{jmh12}$ and $\cite{jmh11}$. In 2010, S. K. Chaubey and R. H. Ojha investigated the M-projective curvature tensor of a Kenmotsu manifold $\cite{jmh13}$.\\\\
$~~~~$The concept of perfect fluid spacetime arose while discussing the structure of this universe. Perfect fluids are often used in the general relativity to model the idealised distribution of matter, such as the interior of a star or isotropic pressure. The energy-momentum tensor $\tilde T$ of a perfect fluid spacetime is given by the following equation 
\begin{equation}
\tilde T(U,V)=pg(U,V)+(\sigma+p)A(U)A(V),
\end{equation}
where $\sigma$ is the energy-density and $p$ is the isotropic pressure, $A$ is defined earlier and the unit vector field $\xi$ is timelike, i.e. $g(\xi, \xi) = - 1$. Einstein field equation with cosmological constant($\cite{jmh30}$) is given by 
\begin{equation}
S(U,V)-\frac{\tilde r}{2}g(U,V)+\lambda g(U,V)=k\tilde T(U,V),
\end{equation}
where, $S$ is the Ricci tensor, $\tilde r$ is the scalar curvature of the spacetime while $\lambda$, $k$ are the cosmological constant and the gravitational constant respectively. It’s used to describe the dark energy of this universe in modern cosmology, which is responsible for the possible acceleration of this universe. The equations $(1.3)$ and $(1.4)$ together give 
\begin{equation}
S(U,V)=(\frac{\tilde r}{2}-\lambda+pk)g(U,V)+k(\sigma+p)A(U)A(V).
\end{equation}
Comparing to the equation $(1.1)$ we can say the tensor of the equation $(1.5)$ represents the tensor of a quasi-Einstein manifold.\\\\
$~~~~$The $k$-nullity distribution $N(k)$ of a Riemannian manifold $M$ is defined by
\begin{eqnarray}
&&N(k):p \rightarrow N_{p}(k)=\nonumber\\
&&\{W\in T_{p}(M):R(U,V)W=k[g(V,W)U-g(U,W)V]\},
\end{eqnarray}
for all $U, V \in T_{p}M$, where $k$ is a smooth function. For a quasi-Einstein manifold $M$, if the generator $\xi$ belongs to some $N(k)$, then $M$ is said to be $N(k)$-quasi-Einstein manifold $\cite{jmh24}$ . \"{O}zg\"
{u}r and Tripathi proved that for an $n$-dimensional $N(k)$-quasi Einstein manifold $\cite{jmh25}$, $k=\frac{l+m}{n-1}$, where $l$ and $m$ are the respective scalar functions and $n$ is the dimension of the manifold.\\\\
$~~~~$In this paper we have first derived some theorems on $M$-projectively flat spacetimes. After that we have introduced the concept of Ricci semi-symmetric spacetime with vanishing $M$-projective curvature tensor. Lastly we introduced the concept of torse-forming vector field in this spacetime and derived some theorems on it, thereby finding the curvature of the spacetime and finding the square of the length of the Ricci tensor for this spacetime with torse-forming vector field.\\

\section{Preliminaries}

$~~~$Consider a quasi-Einstein spacetime with associated scalars $l$, $m$ and associated $1$-form $A$. Then by $(1.1)$, we have 
\begin{equation}
r=4l-m,
\end{equation}
where $r$ is a scalar curvature of the spacetime. If $\xi$ is a unit timelike vector field, then $g(\xi, \xi) = -1$. Again from the equation $(1.1)$, we have
\begin{equation}
S(\xi, \xi)=m-l,
\end{equation}
$~~~~$For all vector fields $U$ and $V$ we have the following equation,
\begin{equation}
g(QU,V)=S(U,V),
\end{equation}
where $Q$ is the symmetric endomorphism of the tangent space at each point of the manifold corresponding to the Ricci tensor $S$. From the equation $(1.5)$ and $(2.3)$ we can get
\begin{equation}
QU=(\frac{\tilde r}{2}-\lambda+pk)U+k(\sigma+p)A(U)\xi.
\end{equation}\\\\
$~~~~$If the unit timelike vector field $\xi$ is a torse-forming vector field ($\cite{jmh15}$, $\cite{jmh14}$) then it satisfies the following equation,
\begin{equation}
\nabla_{U}\xi=U+A(U)\xi.
\end{equation}
In $\cite{jmh9}$ Venkatesha and H. A. Kumara proved that:\\\\
\textbf{Theorem 2.1:} {\em On a perfect fluid spacetime with torse-forming vector field $\xi$, the following relation
holds
\begin{equation}
(\nabla_{U}A)(V)=g(U,V)+A(U)A(V).
\end{equation}}\\\\
$~~~~$Considering a frame field and taking a contraction over $U$ and $V$ from the equation $(1.5)$ we get,
\begin{equation}
\tilde r=4\lambda+k(\sigma-3p).
\end{equation}\\

\section{$M$-projectively flat quasi-Einstein spacetime}

$~~~~~$In this section we consider a quasi-Einstein spacetime with vanishing $M$-projective curvature tensor. If a spacetime with dimension $n = 4$ is $M$-projectively flat then from the equation $(1.2)$ we have
\begin{equation}
R(U,V)W=\frac{1}{6}[S(V,W)U-S(U,W)V+g(V,W)QU-g(U,W)QV].
\end{equation}
Using the equation $(2.3)$ in the equation $(1.1)$ we get
\begin{equation}
QU=lU+mA(U)\xi.
\end{equation}
Using the equations (1.1), (3.2) and taking the inner product with $T$ from $(3.1)$ we get
\begin{eqnarray}
& &\tilde R(U,V,W,T)=\frac{l}{3}[g(V,W)g(U,T)-g(U,W)g(V,T)] \nonumber\\
& &~~~~~~~~~~~~~~~~~+\frac{m}{6}[g(U,T)A(V)A(W)+g(V,W)A(U)A(T)\nonumber\\
& &~~~~~~~~~~~~~~~~~-g(V,T)A(U)A(W)-g(U,W)A(V)A(T)].
\end{eqnarray}
Now taking
\begin{equation}
D(U,V)=\sqrt{\frac{l}{3}}g(U,V)+\frac{m}{2\sqrt{3l}}A(U)A(V),
\end{equation}
from the equation $(3.3)$ we have
\begin{equation}
\tilde R(U,V,W,T)=D(V,W)D(U,T)-D(U,W)D(V,T).
\end{equation}\\
It is known that an $n$-dimensional Riemannian or semi-Riemannian manifold whose curvature tensor $\tilde R$ of type $(0, 4)$ satisfies the condition $(3.5)$, is called a special manifold with the associated symmetric tensor $D$ and is denoted by the symbol  $\psi(D)_{n}$, where $D$ is a symmetric tensor field of type $(0, 2)$. Recently, these types of manifolds are studied in $\cite{jmh16}$ and $\cite{jmh21}$. With the use of the equations (3.4) and (3.5) we can state the following theorem:\\\\
\textbf{Theorem 3.1:} {\em An $M$-projectively flat quasi-Einstein spacetime is $\psi(D)_{4}$, where $D$ is the associated symmetric tensor field.}\\\\
$~~~~$In  $\cite{jmh17}$ B. Y.  Chen and K. Yano introduced the concept of quasi-constant curvature. A manifold is said to be a manifold of quasi-constant curvature if it satisfies the following condition
\begin{eqnarray}
&&\tilde R(U,V,W,T)=p[g(V,W)g(U,T)-g(U,W)g(V,T)]\nonumber\\
&&~~~~~~~~~~~~~~~~~~+q[g(U, T)\eta(V)\eta(W) - g(V, T)\eta(U)\eta(W)\nonumber\\
&&~~~~~~~~~~~~~~~~~~+ g(V, W)\eta(U)\eta(T) - g(U, W)\eta(V\eta(T)],
\end{eqnarray}\\
where $\tilde R$ is the scalar curvature of type (0, 4), $p$ and $q$ are scalar functions while $g(U,\nu) = \eta(U)$, $\nu$ is the unit vector field, $\eta$ is the respective $1$-form and $g(\nu, \nu)=1$. Thus in the view of $(3.3)$ and $(3.6)$ we state the following theorem:\\\\
\textbf{Theorem 3.2:} {\em A spacetime with vanishing $M$-projective curvature tensor is a spacetime of quasi-constant curvature.}\\\\
$~~~~$Now, let us consider the space $\xi^{\perp}=\{X : g(X,\xi) = 0, \forall~X \in \rchi(M)\}$. Let $U$, $V$, $W$ $\in$ $\xi^{\perp}$, then the equation $(3.3)$ will imply
\begin{equation}
R(U,V)W=\frac{l}{3}[g(V,W)U-g(U,W)V].
\end{equation}\\
So, we can state the following theorem:\\\\
\textbf{Theorem 3.3:} {\em An $M$-projectively flat quasi-Einstein spacetime becomes an $N(\frac{l}{3})$-quasi Einstein spacetime provided $U$, $V$, $W$ $\in$ $\xi^{\perp}$, $\xi$ is a unit timelike vector field and $l$ is a non-zero real-valued scalar function.}\\\\
We also derive the following corollary:\\\\
\textbf{Corollary 3.1:} {\em An $M$-projectively flat quasi-Einstein spacetime satisfies the following results,
\begin{eqnarray}
&&~~~~(i) R(U, \xi)V=\frac{m-2l}{6}g(U,V)\xi,\nonumber\\
&& ~~~~(ii) R(U,\xi)\xi=\frac{m-2l}{6}U,\nonumber\\
\end{eqnarray}}
where $U$, $V$ $\in$ $\xi^{\perp}$, $\xi$ is a unit timelike vector field and $l$, $m$ are two non-zero real-valued scalar functions.\\\\
$~~~$After Riemannian manifolds, Lorentzian manifolds form the most important subclass of pseudo-Riemannian manifolds. They are important in applications of general relativity. A principal premise of general relativity is that spacetime can be modelled as a $4$-dimensional Lorentzian manifold of signature $(3,1)$ or, equivalently, $(1,3)$. With a signature of $(3,1)$ or $(1,3)$, the manifold is also time-orientable. A Lorentzian manifold is called infinitesimally spatially isotropic $(\cite{jmh6})$ relative to a unit timelike vector field $\xi$ if its curvature tensor $R$ satisfies the relation
\begin{equation}
R(X,Y)Z=\alpha [g(Y,Z)X-g(X,Z)Y],
\end{equation}
for all $X$, $Y$, $Z$ $\in$ $\xi^{\perp}$ and $R(X,\xi)\xi=\beta X$ for all $X \in \xi^{\perp}$, $\alpha$ and $\beta$ are two non-zero real-valued functions.\\\\
From the equation $(3.7)$ and the result $(ii)$ of corollary $(3.1)$ it is obvious that the manifold is infinitesimally spatially isotropic. Thus we can state the following theorem:\\\\
\textbf{Theorem 3.4:} {\em An $M$-projectively flat quasi-Einstein spacetime is infinitesimally spatially isotropic relative to the unit timelike vector field $\xi$.}\\\\

\section{$M$-projectively flat Ricci semi-symmetric quasi-Einstein spacetime}

$~~~$In this section we consider a quasi-Einstein spacetime which is Ricci semi-symmetric. An $n$-dimensional semi-Riemannian manifold is said to be Ricci semi-symmetric if the tensor $R.S$ and the Tachibana tensor $\tilde Q(g,S)$ are linearly dependent, i.e., 
\begin{equation}
R(U,V) \cdot S(W,T)=F_{S}\tilde Q(g,S)(W,T;U,V)
\end{equation} 

holds on $U_{S}$ where $U_{S}= \{ x \in M : S \neq \frac{r}{n}g~at~x \}$ and $F_{S}$ is a scalar function on $U_{S}$. Now we know that 
\begin{equation}
R(U,V) \cdot S(W,T)=-S(R(U,V)W,T)-S(W,R(U,V)T),
\end{equation}
using the equation $(4.1)$ we have 
\begin{equation}
F_{S}\tilde Q(g,S)(W,T;U,V)=-S(R(U,V)W,T)-S(W,R(U,V)T).
\end{equation} 
We also know that
\begin{equation}
(U \wedge_{g}V)W=g(V,W)U-g(U,W)V.
\end{equation}
Now if it is a Ricci semi-symmetric quasi-Einstein spacetime then using the equations $(4.2)$, $(4.3)$ and $(4.4)$ we get
\begin{eqnarray}
&&S(R(U,V)W,T)+S(W,R(U,V)T)=F_{S}[g(V,W)S(U,T)-g(U,W)S(V,T)\nonumber\\
&&~~~~~~~~~~~~~~~~~~~~~~~~~~~~~~~~~~~~~~~~~~~~~~+g(V,T)S(W,U)-g(U,T)S(V,W)].
\end{eqnarray}
Since we know $\tilde R(U,V,W,T)=-\tilde R(U,V,T,W)$, thus using the equation $(1.1)$ in the equation $(4.5)$ we obtain
\begin{eqnarray}
&&A(R(U,V)W)A(T)+A(W)A(R(U,V)T)\nonumber\\
&&=F_{S}[g(V,W)A(U)A(T)-g(U,W)A(V)A(T)\nonumber\\
&&+g(V,T)A(W)A(U)-g(U,T)A(V)A(W)],
\end{eqnarray}
putting $T=\xi$ in the equation $(4.6)$ and applying the result $g(R(U,V)\xi,\xi)=g(R(\xi,\xi)U,V)$ we get,
\begin{equation}
A(R(U,V)W)=F_{S}[g(V,W)A(U)-g(U,W)A(V)],
\end{equation}
applying the equation $(3.3)$ from $(4.7)$ we get,
\begin{equation}
(F_{S}-\frac{2l-m}{6})[g(V,W)A(U)-g(U,W)A(V)]=0.
\end{equation}
So, if $g(V,W)A(U)-g(U,W)A(V) \neq 0$ then 
\begin{equation}
F_{S}=\frac{2l-m}{6},
\end{equation}
thus using the equations $(4.7)$ and $(4.9)$ we get,
\begin{equation}
R(U,V)W=\frac{2l-m}{6}[g(V,W)U-g(U,W)V],
\end{equation}
from the equations $(1.6)$ and $(4.10)$ we observe that the spacetime becomes an $N(\frac{2l-m}{6})$-quasi Einstein spacetime provided $g(V,W)A(U)-g(U,W)A(V) \neq 0$. This leads us to the next theorem:\\\\
\textbf{Theorem 4.1:} {\em An $M$-projectively flat Ricci semi-symmetric quasi-Einstein spacetime with $g(V,W)A(U)-g(U,W)A(V) \neq 0$ is an $N(\frac{2l-m}{6})$-quasi Einstein spacetime, where $l$ and $m$ are two non-zero real valued scalar functions.}\\
\begin{center}
\section{$M$-projectively flat perfect fluid spacetime with torse-forming vector field}
\end{center}
$~~~~$If a manifold is $M$-projectively flat then using the divergence $\nabla$ to both the sides of the equation $(3.1)$ we get
\begin{equation}
(\nabla_{U}S)(V,W)-(\nabla_{V}S)(U,W)=0,
\end{equation}
using the equation $(2.3)$ we get,
\begin{equation}
g((\nabla_{U}Q)V-(\nabla_{V}Q)U, W)=0.
\end{equation}
From the equation $(2.7)$ since we observe $\tilde r$ is a constant thus using the equation $(2.4)$ we get
\begin{equation}
k(\sigma +p)[(\nabla_{U}A)(V)\xi+A(V)\nabla_{U}\xi-(\nabla_{V}A)(U)\xi-A(U)\nabla_{V}\xi]=0,
\end{equation}
using the equations $(2.5)$ and $(2.6)$ we get
\begin{equation}
k(\sigma+p)[g(V,\xi)U-g(U,\xi)V]=0,
\end{equation}
since $k$ is the gravitational constant hence $k \neq 0$. Thus $g(V,\xi)U-g(U,\xi)V \neq 0$ implies
\begin{equation}
\sigma+p=0,
\end{equation}
which means either $\sigma = p = 0$ (empty spacetime) or the perfect fluid satisfies the vacuum-like equation of state. This allows us to derive the following theorem:\\\\
\textbf{Theorem 5.1:} {\em An $M$-projectively flat perfect fluid spacetime with torse-forming vector field $\xi$ satisfying Einstein field equation with cosmological constant is either an empty spacetime or satisfies the vacuumlike equation of state, provided $g(V,\xi)U-g(U,\xi)V \neq 0$.}\\\\
$~~~~$Now $\sigma+p=0$ means the fluid behaves as a cosmological constant $\cite{jmh18}$. This is also termed as \textbf{Phantom Barrier} $\cite{jmh19}$. Now in cosmology we know such a choice $\sigma=-p$ leads to rapid expansion of the spacetime which is now termed as inflation $\cite{jmh20}$. In $1981$, to explain the conditions observed in the universe, astrophysicist Alan Guth proposed \textbf{cosmic inflation} \cite{jmh26}. The term inflation refers to the explosively rapid expansion of spacetime that occurred a tiny fraction of a second after the \textbf{Big Bang}. So, we obtain the following theorem:\\\\
\textbf{Theorem 5.2:} {\em an $M$-projectively flat perfect fluid spacetime with torse-forming vector field $\xi$ satisfying Einstein field equation with cosmological constant represents an inflation.}\\\\
$~~~~$Now putting $\sigma+p=0$ from the equation $(1.5)$ we get,
\begin{equation}
S(U,V)=[\lambda+\frac{k}{2}(\sigma-p)]g(U,V),
\end{equation}
thus the equation $(2.4)$ becomes
\begin{equation}
QU=[\lambda+\frac{k}{2}(\sigma-p)]U.
\end{equation}
Using the equations $(5.6)$ and $(5.7)$ in the equation $(3.1)$ we get
\begin{equation}
R(U,V)W=\{\frac{2\lambda+k(\sigma-p)}{6}\}[g(V,W)U-g(U,W)V].
\end{equation}
Hence we can state the following theorem:\\\\
\textbf{Theorem 5.3:} {\em An $M$-Projectively flat perfect fluid spacetime with torse-forming vector field $\xi$, satisfying Einstein field equation with cosmological constant is of constant curvature  $\frac{2\lambda+k(\sigma-p)}{6}$.}\\\\
Consequently we obtain the following theorem as:\\\\
\textbf{Theorem 5.4:} {\em An $M$-projectively flat perfect fluid spacetime with torse-forming vector field $\xi$ satisfying Einstein field equation with cosmological constant is an Einstein spacetime.}\\\\
$~~~~$From the equation $(5.8)$ we easily obtain
\begin{eqnarray}
& &(R(U,V) \cdot \tilde R)(X,Y,Z,W)=-\tilde R(R(U,V)X,Y,Z,W)-\tilde R(X,R(U,V)Y,Z,W)\nonumber\\
& &-\tilde R(X,Y,R(U,V)Z,W)-\tilde R(X,Y,Z,R(U,V)W)=0,
\end{eqnarray}
which implies the manifold is semi-symmetric. Hence we obtain the following theorem:\\\\
\textbf{Theorem 5.5:} {\em An $M$-projectively flat perfect fluid spacetime with torse-forming vector field $\xi$ satisfying Einstein field equation with cosmological constant is a semi-symmetric spacetime.}\\\\
$~~~~$Replacing $U$ by $QU$ from the equation $(5.6)$ we get
\begin{equation}
S(QU,V)=[\lambda+\frac{k}{2}(\sigma-p)]g(QU,V).
\end{equation}
Using the equation $(5.6)$which becomes
\begin{equation}
S(QU,V)=[\lambda+\frac{k}{2}(\sigma-p)]S(U,V)=[\lambda+\frac{k}{2}(\sigma-p)]^{2}g(U,V).
\end{equation}
Considering a frame field and taking a contraction over $U$ and $V$ from the equation $(5.11)$ we get
\begin{equation}
||Q||^{2}=4[\lambda+\frac{k}{2}(\sigma-p)]^{2}=[2\lambda+k(\sigma-p)]^{2}.
\end{equation}
Hence we can state the following theorem:\\\\
\textbf{Theorem 5.6:} {\em The square of the length of the Ricci tensor of an $M$-projectively flat perfect fluid spacetime with torse-forming vector field $\xi$ satisfying Einstein field equation with cosmological constant is $[2\lambda+k(\sigma-p)]^{2}$.}\\\\
$~~~~$The Ricci identity is given by
\begin{equation}
\nabla^{2}_{U,V} \alpha (X,Y)-\nabla^{2}_{V,U} \alpha (X,Y)= \alpha (R(U,V)X,Y)+ \alpha (X,R(U,V)Y),
\end{equation}
where $\alpha$ is a symmetric $(0,2)$ tensor. Now if $\alpha$ is parallel to $\nabla$ then
\begin{equation}
\nabla \alpha=0,
\end{equation}
which further implies
\begin{equation}
\nabla^{2} \alpha=0.
\end{equation}
Thus from the equation $(5.13)$ we get
\begin{equation}
\alpha(R(U,V)X,Y)+\alpha(X,R(U,V)Y)=0.
\end{equation}
Thus from the equation $(5.8)$ we get
\begin{eqnarray}
&&\{\frac{2\lambda+k(\sigma-p)}{6}\}[g(V,X)\alpha(U,Y)-g(U,X)\alpha(V,Y)\nonumber\\
&&+g(V,Y)\alpha(U,X)-g(U,Y)\alpha(V,X)]=0.
\end{eqnarray}
Putting $X=Y=V=\xi$ in the equation $(5.17)$ we get,
\begin{equation}
-\{\frac{2\lambda+k(\sigma-p)}{3}\}[\alpha(U,\xi)+A(U)\alpha(\xi,\xi)]=0,
\end{equation}
which means either $\lambda=\frac{k}{2}(p-\sigma)$ or
\begin{equation}
\alpha(U,\xi)=-A(U)\alpha(\xi, \xi).
\end{equation}
Now taking the derivative of $\alpha (\xi,\xi)$ with respect to $V$ and using the equations $(2.5)$ and $(5.19)$ we get
\begin{equation}
V(\alpha(\xi,\xi))=0.
\end{equation}
Taking the derivative of the equation $(5.19)$ with respect to $V$ and using the equation $(5.20)$ we get
\begin{equation}
V(\alpha (U,\xi))=-\alpha(\xi, \xi)V(g(U,\xi)).
\end{equation}
Since $\alpha$ is parallel with respect to $\nabla$ thus using the equation $(2.5)$ from the equation $(5.21)$ we get 
\begin{equation}
\alpha(U,V)=-\alpha(\xi, \xi)g(U,V).
\end{equation}
Therefore we obtain the following theorem as:\\\\
\textbf{Theorem 5.7:} {\em If an $M$-projectively flat perfect fluid spacetime satisfying Einstein field equation with cosmological constant, with torse-forming vector field $\xi$ admits a symmetric $(0,2)$ tensor $\alpha$ parallel to $\nabla$ then either $\lambda = \frac{k}{2}(p-\sigma)$ or $\alpha$ is a constant multiple of $g$.}\\\\

\end{document}